# Coherent Frequency Comb Generation in a Silicon Nitride Microresonator with Anomalous Dispersion

Pei-Hsun Wang[1,*], Yi Xuan[1,2], Jian Wang[1,2], Xiaoxiao Xue[1], Daniel E. Leaird[1], Minghao Qi[1,2], and Andrew M. Weiner[1,2]

[1]School of Electrical and Computer Engineering, Purdue University, 465 Northwestern Avenue, West Lafayette, IN 47907-2035, USA
[2] Birck Nanotechnology Center, Purdue University, 1205 West State Street, West Lafayette, Indiana 47907, USA
*wang1173@purdue.edu

**Abstract:** We observe a transition to a coherent-comb state in a SiN-microresonator with anomalous dispersion. Although ~300 fs pulse trains are generated after line-by-line shaping, the intensity within the microring does not appear to be pulse-like.

Optical frequency comb generation based on high quality factor (Q) microresonators has been widely investigated for its potential in optical communication, optical clocks and pulse generation [1-2]. The correlation between temporal coherence, intensity noise, and communication performance of the generated combs has been well established [3]. Pulse shaping experiments shed light on the time-domain behavior and especially the coherence of generated combs [3-5]. In our previous studies in the normal dispersion regime [3,5], combs that showed high coherence and low intensity noise were observed in some experiments, leading to trains of bandwidth-limited pulses after line-by-line pulse shaping. However, for most of the combs spaced by single free spectral range (FSR), large intensity noise is observed [3,6], accompanied by poor temporal coherence. Some recent studies show the possibility to obtain passive mode-locking of the comb consistent with soliton-pulse formation [7-9] for anomalous cavity dispersion. In this paper, using an anomalous dispersion silicon nitride ring resonator, we observe a transition to a low noise, coherent optical frequency comb state. However, unlike the soliton-pulse formation observed in other experiments [7,8], our comb does not appear to form a pulse directly in the microresonator; subsequent pulse shaping is essential to obtain a clean pulse train.

In the experiment, a silicon nitride ring resonator with 3 µm × 800 nm waveguide cross-section and 100 µm ring radius was fabricated. The gap between the resonator and the bus waveguide is 500 nm, which is close to critical coupling. Figure 1(a) shows the transmission spectrum of a quasi-TE mode, in which the electric field is predominantly polarized parallel to the plane of the resonator. The loaded quality factor (Q) is around $1.1\times 10^6$ at ≈1558.17 nm. The simulated dispersion parameter is 15 ps/nm/km in the anomalous dispersion regime.

The comb dynamics are strongly dependent on the frequency detuning [7,8]. We simultaneously investigate the intensity noise and temporal coherence (pulse compressibility) as the detuning is changed with pump power fixed. The output from the chip is sent into an erbium doped fiber amplifier (EDFA), a combination of standard single-mode fiber and dispersion compensating fiber (DCF), and a pulse shaper. We carefully trim the DCF to achieve dispersion compensation for the entire propagation path subsequent to the SiN chip (with the pulse shaper in a quiescent state). We test this by injecting a pulse from a mode-locked fiber laser into the fiber link directly after the chip and performing intensity autocorrelation based on second harmonic generation with a non-collinear geometry. The measured intensity autocorrelation trace is shown in Fig. 1(b) with a full-width at half-maximum (FWHM) around 288 fs, close to the ideal value of 230 fs (assuming flat phase). Although some evidence of higher order dispersion is seen, we may conclude that the fiber link is dispersion compensated well enough for pulse durations down to several hundred femtoseconds. Therefore, in our subsequent measurements we are able to infer information about the temporal profile of the comb field in the microresonator.

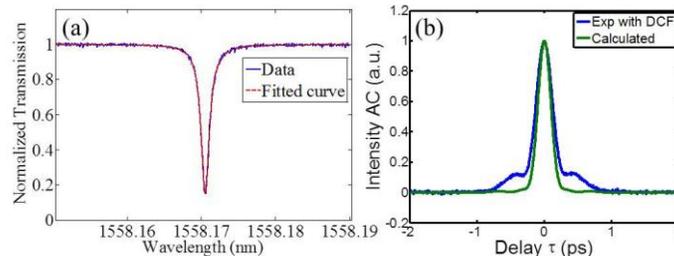

Fig. 1. (a) Transmission spectrum of a fundamental quasi-TE mode with loaded Q≈ 1.1 million. (b) System compensation with DCF fiber. Blue trace shows the measured autocorrelation data while the green trace shows the simulated curve.

Figure 2(a) shows the comb spectrum directly after the chip with input power of 200 mW at around 1558.336 nm. By increasing the input power and slowly tuning the wavelength into the resonance from the short wavelength side, the comb is first generated with 6-FSR spacing once the threshold power (≈40 mW) is reached; and then additional intermediate lines fill in to form a single-FSR comb characterized by high intensity noise – see RF spectrum shown in Fig 2(c). The autocorrelation traces of the generated comb are plotted in Fig. 2(e). The amplitude of the much stronger pump line at the through-port spectrum is attenuated by 10 dB by the programmable pulse shaper. Green and blue lines show the measured autocorrelation traces before and after line-by-line pulse compression. The ideal autocorrelation trace, shown as the red curve in Fig. 2(e), is computed by taking the optical spectrum after the EDFA and assuming flat spectral phase. Clearly, this comb exhibits poor compressibility and degraded coherence. When the pump is further tuned into resonance by 2 pm, we observed a sudden drop in the intensity noise without a dramatic change in the comb spectrum (Fig. 2(b)). In this state line-by-line phase correction successfully compresses the comb into well-formed pulses, in good agreement with the ideal autocorrelation based on stable, flat phase (Fig. 2(f)). The measured autocorrelation peaks have a width of 480 fs FWHM (the intensity duration is estimated as ~300 fs FWHM) and a period of 4.44 ps, corresponding to a 225 GHz repetition rate. Although the comb at this stage exhibits full compressibility and evidently high coherence, the fact that line-by-line shaping is needed to obtain approximately bandwidth-limited pulses implies that the field within the microring is not pulse-like. In this coherent comb state, we are able to retrieve the field information inside the ring based on the optical spectrum and the complement of the phase applied by the shaper to achieve compression. The calculated intensity internal to the ring is shown in Fig. 2(g), while the corresponding phase of the field is shown in Fig. 2(h). We note here that the background level of the intensity in the microring is still uncertain, since it's difficult to know in a through-port device how much of the strong pump line observed at the output port is present inside the resonator. Our results indicating a complexly structured internal comb field are in strong contrast to the observations reported in [7-8], where measurements suggested the presence of well-formed pulses directly in the microresonator.

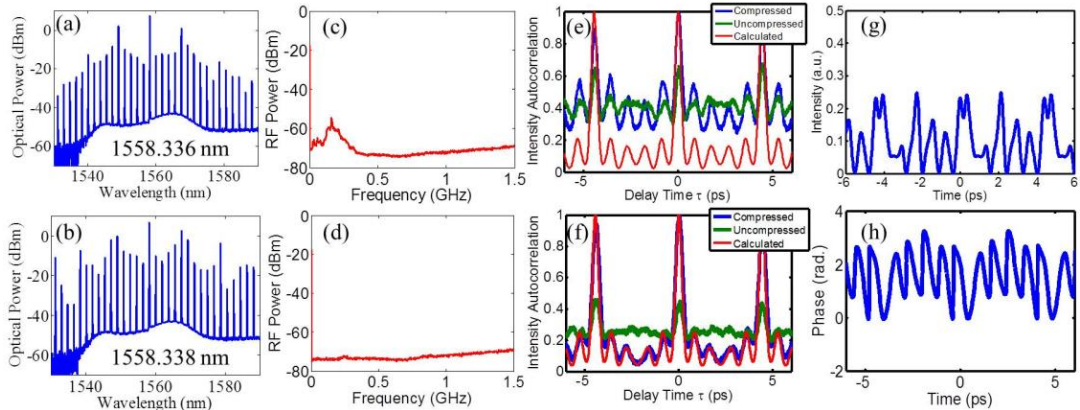

Fig. 2(a) (b) Optical spectra of the generated comb at 1558.336 nm and 1558.338 nm under 200 mW input power. (c) (d) RF spectrum of the generated frequency combs in (a) and (b), respectively. (e) (f) Autocorrelation traces corresponding to the frequency combs generated in (a) and (b). Green and blue lines show the measured autocorrelation trace before and after line-by-line pulse compression while red line shows the intensity autocorrelation trace calculated by taking the optical spectra after the erbium doped fiber amplifier (EDFA) and assuming flat spectral phase. (g) The intensity inside the microring calculated by taking the optical spectrum and the complement of the phase applied to the shaper to realize compression. (h) The corresponding phase of the internal field.

In conclusion, we have demonstrated a transition to high temporal coherence in a silicon nitride microresonator designed for anomalous dispersion. Unlike the soliton formation demonstrated in other reports [7-8], our comb does not appear to form short pulses directly in the cavity. These results suggest the existence of multiple nonlinear operating modes resulting in coherent comb generation.